\documentclass[pre,superscriptaddress,twocolumn,nofootinbib]{revtex4-2}

\usepackage{amsmath,amssymb}
\usepackage{graphicx}
\usepackage[breaklinks=true,colorlinks=true,linkcolor=blue,urlcolor=blue,citecolor=blue]{hyperref}
\usepackage{bm}
\usepackage{comment}
\usepackage{mathrsfs}

\newcommand{\enquote}[1]{\texttt{"}#1\texttt{"}}

\def\O{\mathcal{O}}
\def\arccosh{\mathrm{arccosh}}
\def\tmin{t_{\min}}
\def\Dunit{\mu \mathrm{m}^2/\mathrm{s}}
\def\T{\mathcal{T}}
\def\erf{\mathrm{erf}}

\begin{document}

\title{Fastest first-passage time for multiple searchers with finite speed}

\author{Denis S. Grebenkov}
\affiliation{Laboratoire de Physique de la Mati{\`e}re Condens{\'e}e (UMR
7643), CNRS -- Ecole Polytechnique, IP Paris, 91120 Palaiseau, France}
\author{Ralf Metzler}
\affiliation{Institute of Physics \& Astronomy, University of Potsdam, 14476
Potsdam-Golm, Germany}
\author{Gleb Oshanin}
\affiliation{Sorbonne Universit{\'e}, CNRS, Laboratoire de Physique Th{\'e}orique
de la Mati{\`e}re Condens{\'e}e (UMR CNRS 7600), 4 Place Jussieu, 75252 Paris
Cedex 05, France}

\begin{abstract}
We study analytically and numerically the mean fastest first-passage
time (fFPT) to an immobile target for an ensemble of $N$ independent
finite-speed random searchers driven by dichotomous noise and
described by the telegrapher's equation.  In stark contrast to the
well-studied case of Brownian particles---for which the mean fFPT
vanishes logarithmically with $N$---we uncover that the mean fFPT is
bounded from below by the minimal ballistic travel time, with an
exponentially fast convergence to this bound as $N \to \infty$.  This
behavior reveals a dramatic efficiency advantage of physically
realistic, finite-speed searchers over Brownian ones and illustrates
how diffusive macroscopic models may be conceptually misleading in
predicting the short-time behavior of a physical system.  We extend
our analysis to anomalous diffusion generated by
Riemann-Liouville-type dichotomous noises and find that target
detection is more efficient in the superdiffusive regime, followed by
normal and then subdiffusive regimes, in agreement with physical
intuition and contrary to earlier predictions.
\end{abstract}

\date{\today}


\maketitle

In many biological contexts, diverse agents must efficiently locate
distant targets---such as transcription factors binding to specific
operators on the cellular DNA, immune cells detecting pathogens, or
signaling molecules reaching their conjugate receptors
\cite{ptashne,alberts,olivier}. These processes are inherently
stochastic and often proceed in crowded dynamic environments such as
cellular cytoplasm. Although the first-passage statistics for
individual searchers are quite well understood
\cite{elf,isaacsson,maxplos}, biological systems rarely rely on 
a single agent; instead, multiple searchers are deployed in parallel
to speed up  target detection \cite{target,alberts,snustad}.

When $N$ searchers begin simultaneously, target detection is determined not by
a single agent's shortest arrival time $\tau$, but competitively by the earliest
arrival time among all searchers. The problem thus shifts from individual
first-passage times $\tau_k$, $k=1,2,\ldots,N$, to the statistic of the extremal
random variable $\T_N=\min\{\tau_1,\ldots,\tau_N\}$---the fastest first-passage
time (fFPT). Order statistics therefore play a central role: the key aspects are
the fFPT distribution and how its moments scale with $N$, characterizing the
efficiency of such a multi-agent search. The analysis of such multiple searcher
dynamics reveals how redundancy improves both the reliability and speed of
biological search, offering insight into how living systems may optimize their
performance.

Most existing analyses of multi-agent search assume that all agents
start from the same point simultaneously and move independently as
Brownian walkers with the same diffusion coefficient $D$
\cite{Katja,Meerson15,Reynaud15,Lawley20c,Lawley20d,Lawley20e,Madrid20,Schuss19}. Their
positions follow Langevin dynamics with Gaussian white noise, and the
position probability density function (PDF) of each searcher evolves
according to the diffusion equation from $t=0$, a framework that
accurately captures the long-time behavior of many stochastic
processes in nature. Within this setting, the \textit{mean\/} fFPT to
a target at a distance $x_0$ follows the inverse-logarithmic law
\cite{Katja,Meerson15,Reynaud15,Lawley20c,Lawley20d,Lawley20e,Madrid20,Schuss19}
\begin{equation}
\label{eq:TN_diff}
\overline{\T_N}\simeq\frac{x_0^2}{4D\ln(N)}\quad(N\to\infty). 
\end{equation}
Here the bar denotes averaging with respect to individual trajectory
realizations of all searchers, and the symbol \enquote{$\simeq$}
signifies that we consider solely the leading-order behavior in the
limit $N\to\infty$. This result was first derived for one-dimensional
continuous-space systems, but later shown to hold in bounded domains
of any dimension \cite{Grebenkov2020,adam}, because it is dominated by
so-called \enquote{direct} trajectories \cite{aljazprx,deniscomm}
that go straight to the target, rendering the actual embedding spatial
dimension irrelevant.

The asymptotic form \eqref{eq:TN_diff} shows that deploying more and
more searchers steadily lowers the mean fFPT, albeit only
logarithmically with $N$. In the limit $N\to\infty$, the fastest
searcher would thus reach the distant target arbitrarily
quickly---effectively instantaneously. Moreover, assuming that the
agents undergo a subdiffusive motion with the mean-squared
displacement (MSD) $\overline{x^2(t)}\propto t^{\alpha}$ ($0<
\alpha<1$) and a position PDF obeying a fractional diffusion equation
\cite{pccp}, it was shown that
\begin{equation}
\label{eq:TN_subdiff}
\overline{\T_N}\simeq\frac{t_{\alpha}}{[\ln(N)]^{\frac{2}{\alpha}-1}}\quad
(N\to\infty),
\end{equation}
where $t_{\alpha}$ is a characteristic time-scale
\cite{Lawly2020}. This asymptotic result indicates that for
subdiffusive dynamics, $\overline{\T_N}$ also vanishes as
$N\to\infty$. Strikingly, Eq.~\eqref{eq:TN_subdiff} implies that
$\overline{\T_N}\to 0$ when $\alpha\to 0$, suggesting that
\textit{slower\/} diffusion \textit{enhances\/} the speed of the
fastest arrival, as highlighted in the title of
\cite{Lawly2020}. Although mathematically rigorous, the results
\eqref{eq:TN_diff} and \eqref{eq:TN_subdiff} are clearly unrealistic---even the
fastest searcher needs a finite time to reach a target a finite distance
away---thus underscoring the need for a more refined analysis that yields
more plausible behavior.

Here, we revisit this long-standing problem by assuming that
individual searchers follow a one-dimensional generalized Langevin
dynamics driven by symmetric
\textit{dichotomous\/} noise---a stochastic motion with random switching between
velocities $\pm v$ with rate $\lambda$ \cite{Hanggi95,Bena06} (see
also \cite{Sandev22,pastur} and references therein). This choice
avoids the unphysical behavior inherent in Gaussian white-noise
models, where a searcher has a non-zero probability of appearing
arbitrarily far from its starting point in arbitrarily short time,
allowing for unrealistically small first-passage times (historically,
this problem is well known in heat transport, where it is circumvented
by replacing the parabolic diffusion equation by a hyperbolic Cattaneo
equation with finite propagation speed \cite{cattaneo,jou}). For large
$N$, such short-time artifacts dominate the moments of $\T_N$,
over-estimating the survival probabilities. Dichotomous noise thus
provides a more realistic framework to capture the short-time dynamics
relevant to multi-agent search.  Moreover, symmetric dichotomous noise
with alternating velocities naturally encodes strong
antipersistence---common in crowded environments---since each forward
step is followed by a backward one (see, e.g., \cite{we}). We also
note that this very framework has been successfully used to model
bacterial and other active-particle dynamics, yielding physically
realistic behavior (see, e.g.,
\cite{t2,t3,kurz}). Particularly, the motion of sperm cells, one of the major
biological motivations for studying the mean fFPT \cite{Reynaud15,Schuss19},
would be more naturally described by (biased) dichotomous noises than by white
Gaussian ones.

We show that the dichotomous-noise framework naturally removes the
above unphysical artifacts and provides a more physically realistic
and consistent behavior of the mean fFPT $\overline{\T_N}$.  Within
this picture, $\overline{ \T_N}$ tends to $\tmin=x_0/v$ as
$N\to\infty$---the ballistic travel time and hence the natural lower
bound on $\overline{\T_N}$. The convergence to this value with
increasing $N$ is entirely controlled by the key parameter, the
dimensionless ballistic travel time number $\gamma=x_0\lambda/v$. The
parameter $\gamma$ can be quite small for particles moving in
viscoelastic media such as the cell cytoplasm, or for
bacteria. Conversely, it may attain large values for particles moving
in acqeous solutions. We show below that $\gamma$ determines the
characteristic number $N_\gamma$ that separates two asymptotic
regimes: For $N\gg N_\gamma$, the approach of $\overline{\T_N}$ to
$\tmin$ is described by an exponential function of $N$,
$\overline{\T_N}-\tmin\propto e^{-N/N_\gamma}$, which signifies that
deploying $N$ searchers may actually be, by far, a more efficient
strategy than one might expect from the logarithmic reduction
predicted by Eqs.~\eqref{eq:TN_diff} and \eqref{eq:TN_subdiff}. In
contrast, for $N$ in the interval $3\leq N\ll N_\gamma$, the approach
is quite slow and for $\gamma\to\infty$ (and hence
$N_{\gamma}\to\infty$) becomes consistent with the dependence in
Eq.~\eqref{eq:TN_diff}. Concurrently, the result in
Eq.~\eqref{eq:TN_diff} follows directly from our analysis if we first
take the \enquote{diffusion limit} of the dichotomous noise and then
consider the limit $N\to\infty$, which demonstrates that these limits
do not commute. To probe the behavior of $\overline{\T_N}$ for
\textit{anomalous\/} diffusion, we consider the dynamics driven by the
Riemann-Liouville fractional dichotomous noise
\cite{Dean21}. We show that also in this case the mean fFPT tends to some
minimal travel time (see Eq.~\eqref{eq:tmin_H} below) in the limit
$N\to\infty$ and, for physically relevant parameters, the approach is
faster in the superdiffusive regime than in the diffusive one, and the
latter outperforms subdiffusion, which is an intuitively expected and
correct trend, as compared to the one predicted by
Eq.~\eqref{eq:TN_subdiff}.

\emph{Model}. Consider a one-dimensional system in which $N$ particles are
released from a point $x_0>0$ at time $t=0$ and search for an immobile target
at the origin. All $N$ particles move on the positive halfline independently of
each other, and their instantaneous positions $x_k(t)$, $k=1,2,\ldots,N$, obey
the stochastic differential equations
\begin{equation}
\label{dich}
\dot{x}_k(t)=\eta_k(t),\quad x_k(0)=x_0,
\end{equation}
where $\eta_k(t)$ are statistically-independent \textit{symmetric\/}
dichotomous noises alternating between the values $\pm v$. The switching
events occur after random, exponentially distributed time intervals with
the rate $\lambda$. The mean and autocovariance function of the noises
read \cite{Hanggi95,Bena06}
\begin{equation}
\overline{\eta_k(t)}=0,\quad\overline{\eta_k(t)\eta_{k'}(t')}=\delta_{k,k'}
2\lambda De^{-2\lambda|t-t'|},
\end{equation}
in terms of the Kronecker-delta $\delta_{k,k'}$ and the long-time diffusion
coefficient $D=v^2/(2\lambda)$. Here $1/(2\lambda)$ is a finite correlation
time of the noises. The MSD of $x_k(t)$ scales linearly with time in the
long-time limit. Based on this model we focus on the statistics of the
first-passage times $\tau_k$ to the target of the respective searchers and
on the above-defined extremal random variable $\T_N$, the fFPT.

We also study dichotomous dynamics in a broader context by extending it to
anomalous diffusion---processes characterized by an anomalous diffusion
exponent $\alpha$ that may be less than unity (subdiffusion) or greater than
unity (superdiffusion) \cite{pccp}. We model anomalous diffusion by a
Riemann-Liouville-type fractional dichotomous process as introduced in
\cite{Dean21}, in which individual searcher trajectories are described by
the stochastic integral
\begin{eqnarray}
\label{eq:Langevin} 
x_k(t)&=&x_0+\int_0^tK(t-t')\eta_k(t')dt',\\
\nonumber
K(t)&=& (t/T_0)^{(\alpha-1)/2}/\Gamma((\alpha+1)/2),\quad0<\alpha<2,
\end{eqnarray}
where $T_0$ is a characteristic time-scale of the power-law memory kernel $K(t)$.
The MSD of such a process grows with time as $\overline{x^2(t)}\propto t^{\alpha}$,
in analogy to the fractional dynamics studied in \cite{Lawly2020}. Note that the
centered process $y_k(t)=x_k(t)-x_0$ evolves inside the \enquote{light cone} $|y_k(t)|
\leq y^*(t)$ for any $t$, where $y^*(t)\sim vT_0(t/T_0)^{(1+\alpha)/2}$. The
position PDF of $y_k(t)$, which is supported on $[-y^*(t),y^*(t)]$ and
vanishes outside of this support, was studied in detail in
\cite{Dean21}. The PDF is approximately Gaussian in its central part
and vanishes at the endpoints $\pm y^*(t)$ of the support.  For this
model we will also study the statistics of the first-passage times and
the mean fFPT by extensive numerical simulations.

\begin{figure}
\includegraphics[width=0.99\columnwidth]{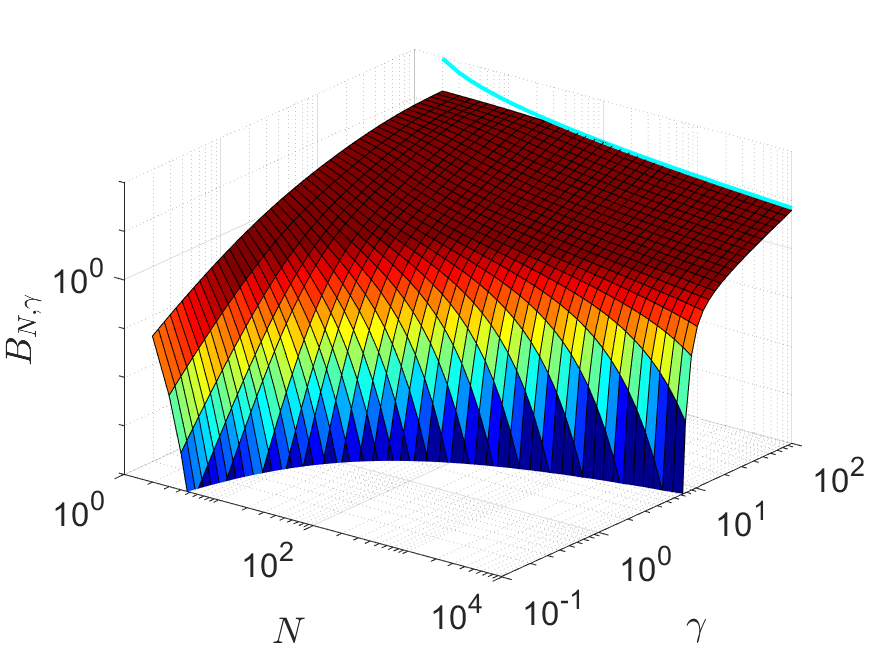} 
\caption{(Color online) 3D plot of $B_{N,\gamma}$ obtained from numerical
evaluation of the integrals in Eqs.~\eqref{eq:B_def} and \eqref{eq:f_def},
as function of $N$ and $\gamma$. The cyan solid line indicates our
asymptotic large-$\gamma$ prediction $2\gamma/(\pi\ln N)$, see
Eqs.~\eqref{inter} and \eqref{z}.}
\label{fig:BN_surf}
\end{figure}

\emph{Central results}. The PDF of first-passage times for a single
particle moving randomly subject to a symmetric dichotomous noise was
analyzed in \cite{Malakar18} (see also \cite{Mori20,urna}).  For
completeness, a derivation is presented in the Supplemental Material
(SM), Sec. A \cite{sm}. Our main goal here is to determine the mean
fFPT $\overline{\T_N}$ and to study its functional dependence on $N$
and $\gamma$. Relegating details of calculations to SM.B1 \cite{sm},
we find that for arbitrary $N\geq3$ and $\gamma>0$,
\begin{equation}
\label{eq:TN_exact}
\overline{\T_N}=t_{\min}\big(1+B_{N,\gamma}\big),\quad t_{\min}=x_0/v,
\end{equation}
where $t_{\min}$ is the ballistic travel time, and the dimensionless function
$B_{N,\gamma}$ quantifies the relative excess induced by the dichotomous
dynamics of $N$ searchers,
\begin{subequations}
\label{Bf}
\begin{eqnarray}
\label{eq:B_def}
B_{N,\gamma}&=&\int_1^{\infty}[f_\gamma(y)]^Ndy,\\
\label{eq:f_def}
f_\gamma(y)&=&\gamma\int_{\arccosh(y)}^{\infty}e^{-\gamma\cosh z}
I_1(\gamma\sinh z)dz,
\end{eqnarray}
\end{subequations}
where $I_1(z)$ is the modified Bessel function of the first kind and $\gamma$
the above-mentioned dimensionless number $\gamma=x_0\lambda/v=v x_0/(2D)$. Note
that $B_{N,\gamma}\geq0$ so that $\overline{\T_N}$ cannot become smaller than
$\tmin$, i.e., $\tmin$ is a natural lower bound on the mean fFPT. Furthermore,
$B_{N,\gamma}$ is a monotonically \textit{increasing\/} function of $\gamma$ and
a monotonically \textit{decreasing\/} function of $N$, as intuitively expected.
In Fig.~\ref{fig:BN_surf} we plot $B_{N,\gamma}$ obtained from numerical
evaluation of the integrals in Eqs.~\eqref{eq:B_def} and \eqref{eq:f_def}, which
illustrates its overall behavior and clearly highlights these features.

Before we discuss the asymptotic behavior of expression \eqref{eq:TN_exact},
we consider an \enquote{order of magnitude} estimate of $\gamma$ for several physical
systems. For instance, the diffusive dynamics of a small protein or of an ion,
say $K^+$, in water, has a noise correlation time of the order of picoseconds,
$(2\lambda)^{-1}\sim 10^{-12}\,$s. The diffusion coefficients of a small
protein or of an ion are of the order of $10^2\,\Dunit$ \cite{protein_water}
and $10^3\,\Dunit$ \cite{ion_water}, respectively, such that $\gamma\sim 10^4
\,x_0/\mu$m or even $\gamma\sim 10^5\,x_0/\mu$m. For $x_0$ of the order of a
few micrometers, $\gamma$ is therefore very large. In turn, in a crowded
environment, e.g., the cellular cytoplasm, the correlation time of noise is
much longer: $(2\lambda)^{-1}$ typically ranges from milliseconds to seconds
\cite{timeref1,timeref2}. Concurrently, the diffusion coefficient of a small
protein in cytoplasm is typically suppressed by more than an order of
magnitude compared to water \cite{protein_cell} and the ionic
diffusion is reduced by a factor of a few
\cite{ion_water}. Respectively, one finds $\gamma\sim(1-10)\,
x_0/\mu$m. Again, for $x_0$ in the range of a few micrometers,
$\gamma$ should attain in this physical situation very modest
values. Lastly, consider the dynamics of bacteria such as {\it
E. coli\/}, which performs paradigmatic run-and-tumble motion
\cite{t2,t3}. A systematic analysis of the experimental data has been
recently performed in \cite{kurz}, giving $D\approx0.3\,\Dunit$,
$v\approx17\,\mu\mathrm{m/s}$ and the mean run time of order of
$3$~s. This leads to $\gamma\approx 0.5\, x_0/\mu$m, and hence, for a
target site placed at a distance of a few micrometers away, $\gamma$
amounts to several units.  Therefore, all values of $\gamma$ are in
principle relevant but correspond to specific physical situations:
while $\gamma$ can be large for the dynamics in aqueous solutions,
moderate or even relatively small values correspond to the motion of
bacteria, or ions and small proteins in cellular cytoplasm.

\begin{figure}
\includegraphics[width=0.99\columnwidth]{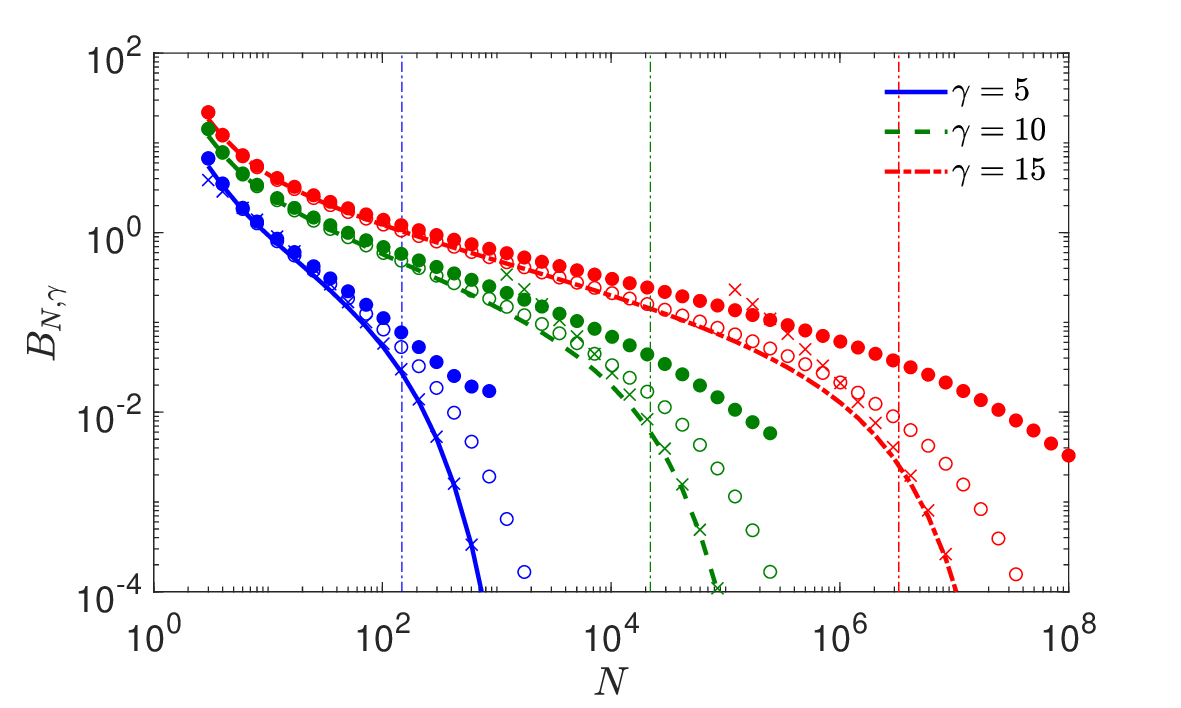} 
\caption{(Color online) Excess factor $B_{N,\gamma}$ for $\gamma=5$ (blue),
$\gamma=10$ (green) and $\gamma=15$ (red) as function of $N$. Colored solid,
dashed and dot-dashed curves depict the corresponding factors $B_{N,\gamma}$
obtained by a numerical evaluation of the integrals in Eqs.~\eqref{eq:B_def}
and \eqref{eq:f_def}. Crosses present the large-$N$ asymptotic form in
Eq.~\eqref{next}. Empty colored circles show the intermediate-$N$ asymptotic
form in Eq.~\eqref{eq:tt0}, whereas the filled circles indicate the asymptotic
relation \eqref{inter}. Colored vertical thin dash-dotted lines represent
$N_\gamma\approx148$, $N_\gamma\approx2.2\times10^4$ and $N_\gamma\approx3.3\times10^6$,
respectively.}
\label{fig:BN}
\end{figure}

\textit{Asymptotic large-$N$ form of $B_{N,\gamma}$}. We first consider the
regime when $\gamma$ is fixed and $N\gg N_{\gamma}$, where 
\begin{equation}
\label{eq:Ngamma}
N_{\gamma}=-1/\ln\left(1-e^{-\gamma}\right),
\end{equation}
a crucial parameter setting the scale for the number of searchers
beyond which the exponential decay becomes relevant.  We show in SM.B2
\cite{sm} that in this regime
\begin{equation}
\label{next}
B_{N,\gamma}\approx2\left(e^{\gamma}-1\right)(\gamma^2N)^{-1}e^{-N/N_{\gamma}}
\quad(N\gg N_{\gamma}),
\end{equation}
which demonstrates that the large-$N$ decay of $B_{N,\gamma}$ is always
\textit{exponential}, in striking contrast with the prediction in
Eq.~\eqref{eq:TN_diff} for Brownian walkers. In Fig.~\ref{fig:BN}, we
depict the asymptotic form in Eq.~\eqref{next} for different values of
$\gamma$ together with the corresponding values of $N_{\gamma}$ (crosses
and vertical dashed lines), alongside with numerical evaluations of the
integrals in Eqs.~\eqref{eq:B_def} and \eqref{eq:f_def} (colored solid,
dashed and dot-dashed curves). We observe excellent agreement between our
prediction in Eq.~\eqref{next} and the numerical evaluation of the excess
factor $B_{N,\gamma}$.

\textit{Intermediate-$N$ form of $B_{N,\gamma}$}. By virtue of
Eq.~\eqref{eq:Ngamma}, the threshold value is $N_{\gamma}\simeq
e^{\gamma}$ for large $\gamma$, so that the range in which the
large-$N$ exponential behavior can be observed shifts to extremely
large (and even unphysical) values of $N$.  Indeed, while $N_{\gamma}$
has quite modest values for small $\gamma$ (e.g., $N_{\gamma}\approx
2.18$ for $\gamma=1$, $N_{\gamma}\approx 6.88$ for $\gamma=2$ and
$N_{\gamma}\approx 148$ for $\gamma=5$), already for $\gamma=26$ it
exceeds $10^{11}$, beyond practically realizable numbers of deployed
searchers. Therefore, the intermediate-$N$ regime $3\leq N\ll
N_{\gamma}$ is a physically relevant domain for systems in which
$\gamma$ is large, e.g., in aqueous solutions.

The asymptotic analysis in this regime is much subtler. In SM.B3 \cite{sm}, we
show that $B_{N,\gamma}$ can be accurately approximated in this $N$-range as
\begin{equation}
\label{eq:tt0}
B_{N,\gamma}\approx\int_0^1\frac{1-x^4}{x^3}\left[\erf\left(\sqrt{\gamma}
x\right)\right]^Ndx,
\end{equation}
where $\erf(x)$ is the error function. In Fig.~\ref{fig:BN} we depict the
approximate form \eqref{eq:tt0} by empty colored circles, confronting this
prediction with the numerically evaluated integrals in Eqs.~\eqref{eq:B_def}
and \eqref{eq:f_def}. We observe perfect agreement over an extended range of
$N$, which broadens progressively as $\gamma$ increases. Furthermore, using
the tight upper bound on $\erf(x)$ \cite{will} and assuming that $N$ is even,
we find the following asymptotic relation which holds uniformly for any $N$
within the interval $3\leq N\ll N_{\gamma}$,
\begin{equation}
\label{inter}
B_{N,\gamma}\approx\frac{2\gamma}{\pi}S_{N/2}-1+\frac{\pi}{8\gamma}H_{N/2}+
\O\left(e^{-\gamma}\right),
\end{equation}
where $H_n=\sum_{j=1}^n 1/j$ is the harmonic number, and
\begin{equation}
\label{eq:Sn}
S_n=\sum_{j=1}^n(-1)^j\binom{n}{j}j\ln(j).
\end{equation} 
The above result is presented in Fig.~\ref{fig:BN} by filled colored
circles and shows that Eq.~\eqref{inter} provides an accurate
approximation of the behavior of the excess factor $B_{N,\gamma}$ for
intermediate values of $N$. In the limit $\gamma\to\infty$, one has
$N_{\gamma}\to\infty$, and we may consider the asymptotic behavior of
$B_{N,\gamma}$ when $N\to\infty$. In this limit the dominant
contribution to $B_{N,\gamma}$ in Eq.~\eqref{inter} comes from the
first term proportional to $\gamma$. Taking into account that (see
SM.B3 \cite{sm})
\begin{equation}
\label{z}
S_{N/2}\simeq\frac{1}{\ln(N/2)}-\frac{C}{\ln^2(N/2)}+\frac{C^2+\pi^2/6}{\ln^3
(N/2)},
\end{equation}
where $C=0.577..$\ is the Euler-Mascheroni constant, and using
Eqs.~\eqref{inter} and \eqref{eq:TN_exact}, we find that, to leading
order in $N$, $\overline{\T_N}\simeq x_0^2/(\pi D\ln N)$. This expression
agrees with Eq.~(\ref{eq:TN_diff}) for Brownian walkers apart from the
numerical factor ($1/\pi$ instead of $1/4$). The function $2\gamma/(\pi
\ln(N))$ is depicted by the thick solid line in Fig.~\ref{fig:BN_surf}
showing a very good agreement with the numerically evaluated
$B_{N,\gamma}$.  For a discussion of how one recovers
Eq.~\eqref{eq:TN_diff} as the diffusion limit of the dichotomous
noise, see SM.B4 \cite{sm}.

\begin{figure}
\includegraphics[width=0.99\columnwidth]{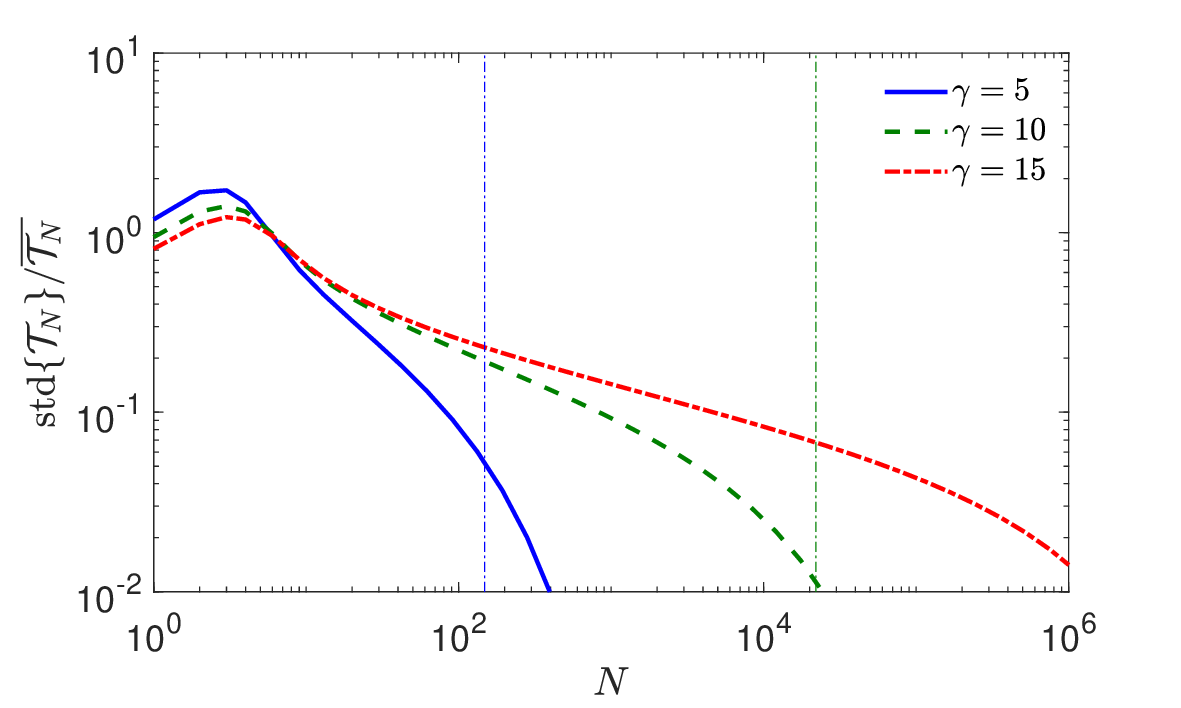} 
\caption{(Color online) Coefficient of variation $\kappa$ from Eq.~\eqref{kappa}
as function of $N$ for three values of $\gamma$.}
\label{fig:std_mean1}
\end{figure}

It is legitimate to ask whether the mean fFPT is representative of the actual
behavior. To provide an answer, we consider the coefficient of variation
$\kappa$ of the fFPT, defined as the ratio of the standard deviation of the
realization-dependent fFPT divided by its mean,
\begin{equation}
\label{kappa}
\kappa=\frac{\sqrt{\overline{\T_N^2}-\overline{\T_N}^2}}{\overline{\T_N}}.
\end{equation}
This quantity is illustrated in Fig.~\ref{fig:std_mean1}, where
$\kappa$ is shown as function of $N$ for three values of $\gamma$. We
find that $\kappa$ exceeds unity only when a small number of searchers
is deployed, which is intuitively expected. For larger $N$, however,
$\kappa$ drops sharply below unity, indicating that fluctuations of
the fFPT become negligible and that the arrival times of the searchers
concentrate around the common value $\tmin$: the mean fFPT provides a
faithful characterization of the search dynamics. In fact, in the
regime $N\gg N_{\gamma}$, we find (see SM.B5 \cite{sm}) that $\kappa$
decays \textit{exponentially\/} with $N$,
\begin{equation}
\kappa\approx\frac{2\sqrt{2}(e^{\gamma}-1)}{\gamma^2N}e^{-N/(2N_{\gamma})}.
\end{equation}

\textit{Anomalous diffusion.} In contrast to Brownian motion, anomalous
diffusion is not universal in the sense that many physically different
stochastic processes give rise to the MSD-scaling $\overline{x^2(t)}\propto
t^{\alpha}$ \cite{pccp,manzo,henrik}. A flexible prototype process is
Riemann-Liouville fractional Brownian motion, a variant of Mandelbrot-van Ness
fractional Brownian motion \cite{fbm,wei,mishura2008} based on fractional
Gaussian noise. Fractional Brownian motion has been shown to provide good
quantitative descriptions of anomalous-diffusive motion in a variety of
viscoelastic media and even for animal motion \cite{vilk} (and references
therein). 

\begin{figure}
\includegraphics[width=0.99\columnwidth]{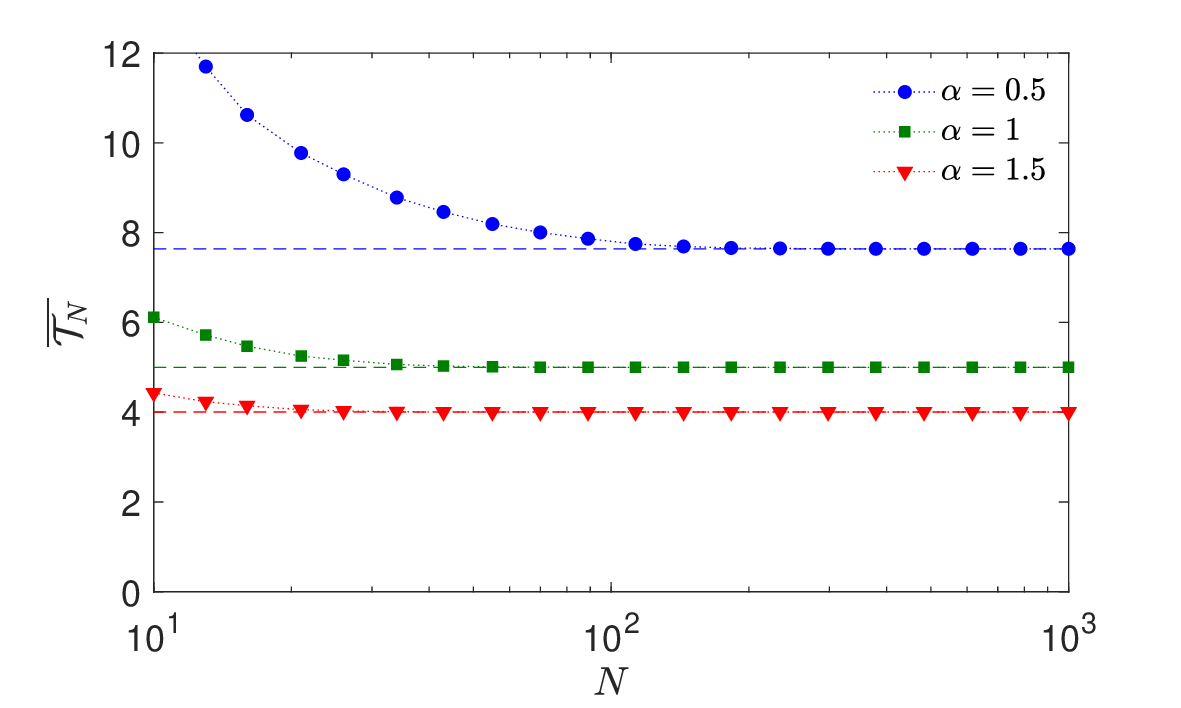} 
\caption{
Mean fFPT $\overline{\T_N}$ as a function of $N$ for the dynamics
generated by Riemann-Liouville dichotomous noise for three values of
the scaling exponent: $\alpha = 0.5$ (subdiffusive behavior, blue),
$\alpha = 1$ (diffusive behavior, green) and $\alpha = 1.5$
(superdiffusive behavior, red). The symbols depict the results of
simulations with $M=10^4$ particles, for $T_0=1$, $D=1$, $v = 1$,
$\lambda=v^2/(2D)$, and $x_0=5$ (such that $x_0/(v T_0) > 1$). The
horizontal dashed lines indicate the corresponding ballistic travels
times in Eq.~\eqref{eq:tmin_H}.}
\label{fig:TN_H1}
\end{figure}

We consider a Riemann-Liouville fractional dichotomous process
\cite{Dean21}, which preserves the essential property of
finite-speed propagation.  This is a convenient unifying framework to
illustrate the consequences for the search efficiency in the
$N$-searcher case. When a typical searcher performs many jumps before
reaching the target and the timescale $T_0$ of the memory kernel is
sufficiently small to ensure $x_0/(vT_0)>1$, we show in SM.C \cite{sm}
that the minimal travel time reads
\begin{equation}
\label{eq:tmin_H}
\tmin = T_0 \biggl(\frac{x_0\Gamma((3+\alpha)/2)}{vT_0}\biggr)^{2/(1+\alpha)},
\end{equation}
which is a \textit{decreasing\/} function of $\alpha$.  In other
words, $\tmin$ is smaller in the superdiffusive regime than in the
diffusive one, and that one is again less than for subdiffusion---a
trend which is apparent in Fig.~\ref{fig:TN_H1}.  Moreover, we observe
that the approach of the mean fFPT to $\tmin$ on an increase of $N$ is
fastest for superdiffusion and longest for subdiffusion, which agrees
with physical intuition. The trend is reversed for $x_0/(vT_0)<1$,
when the target can be reached in a single step (see SM.C \cite{sm}).

\emph{Conclusions.} We revisited the biologically relevant problem of parallel
search by $N$ independent random searchers for an immobile target, a
paradigmatic setting for extreme first-passage statistics in many
applications.  In the standard Brownian-motion framework, the mean
fFPT decreases only logarithmically with $N$ and formally vanishes as
$N\to\infty$, implying both instantaneous arrival in the infinite-$N$
limit and a weak benefit from deploying multiple searchers. This
framework further predicts subdiffusive dynamics to outperform normal
diffusion---an artifact of the unphysical possibility of arbitrarily
fast propagation.

Fixing this shortcoming, we formulated the search dynamics using
dichotomous noise, leading to the telegrapher's equation with finite
propagation speed and compact, time-dependent support of the position
PDF. In this physically consistent setting, the mean fFPT converges,
as $N\to\infty$, to a finite lower bound equal to the shortest
possible, ballistic travel time. Remarkably, the convergence to this
bound is exponentially fast in $N$, demonstrating a dramatic
efficiency gain over Brownian search. From a biological perspective,
it is thus meaningful to invest the energy to produce many searchers
to substantially speed up underlying random search processes. For
intermediate $N$, the convergence remains slow and mimics the familiar
logarithmic behavior, thereby reconciling earlier results with
finite-speed dynamics.

We extended our analysis to anomalous transport via the
Riemann-Liouville dichotomous process. This yielded a clear hierarchy:
superdiffusive search is most efficient, followed by diffusive scaling
and then subdiffusion, in contrast to predictions from
continuous-space fractional diffusion models.

Overall, our results show that imposing a finite propagation speed
fundamentally alters the extreme first-passage behavior, resolves
long-standing paradoxes of Brownian statistics, and identifies
finite-velocity searchers as a superior strategy for rapid parallel
target detection. Future research directions include determining the
full fFPT distribution and extending the discrete-space approach
developed in Ref.~\cite{Lawley20b}.

\begin{acknowledgments}
D.S.G acknowledges partial support by the Alexander von Humboldt Foundation
within a Bessel Prize award. RM acknowledges grant ME 1535/22-1 from the
German Science Foundation (DFG).
\end{acknowledgments}

\clearpage

\renewcommand{\theequation}{S\arabic{equation}}
\renewcommand{\thefigure}{S\arabic{figure}}
\renewcommand{\thesection}{SM.\Alph{section}}
\renewcommand{\thesubsection}{\arabic{subsection}}

\onecolumngrid

\begin{center}
\textbf{\large Supplemental Material:\\[0.32cm]
Fastest first-passage time for multiple searchers with finite speed}\\[0.4cm]
Denis S. Grebenkov, Ralf Metzler, and Gleb Oshanin\\[0.4cm]
In this Supplemental Material (SM), we provide additional details on the exact solution of
the boundary value problem for a single particle, the calculation of the
fastest first-passage time and its detailed properties, as well as on the
simulation of anomalous-diffusion dynamics.
\end{center}

\twocolumngrid

\section{Exact solution for a single particle}
\label{sec:single}

For completeness, we compute the survival probability of a {\it single\/}
particle on the positive semi-axis. This computation was performed in
\cite{Malakar18} (see also \cite{urna,Mori20}). Here we start from the same
backward Fokker-Planck equations and provide an alternative computation,
which yields the same results. Let $S_{\pm}(t|x_0)=\mathbb{P}_{x_0,\pm}\{\tau
>t\}$ denote the survival probability for a particle launched at $x_0$ with
the velocity $\pm v$. The switching between positive and negative velocities
can be easily implemented into the backward Fokker-Planck eqution:
\begin{eqnarray}
\nonumber
\partial_tS_+(t|x_0)&=&v\partial_{x_0}S_+(t|x_0)-\lambda S_+(t|x_0)+\lambda S_-
(t|x_0),\\
\nonumber
\partial_tS_-(t|x_0)&=&-v\partial_{x_0}S_-(t|x_0)-\lambda S_-(t|x_0)+\lambda S_+
(t|x_0).\\
\end{eqnarray}
The action of the time derivative onto the second equation reads
\begin{eqnarray}
\nonumber
\partial_t^2S_-&=&-(\lambda+v\partial_{x_0})\partial_tS_-+\lambda\partial_tS_+\\
\nonumber
&=&-(\lambda+v\partial_{x_0})\bigl[-v\partial_{x_0}S_-+\lambda(S_+-S_-)\bigr]\\
\nonumber
&+&\lambda\bigl[v\partial_{x_0}S_+-\lambda(S_+-S_-)\bigr]\\
&=&v^2\partial_{x_0}^2S_--2\lambda\partial_tS_-,
\end{eqnarray}
i.e., one recovers the standard form of the telegrapher's equation (see, e.g.,
\cite{Bena06}),
\begin{equation}
\label{eq:St_eq}
(\partial_t^2+2\lambda\partial_t-v^2\partial_{x_0}^2)S(t|x_0)=0\qquad(x_0>0),
\end{equation}
where we dropped the subscript minus. The initial condition for the
survival probability is $S(0|x_0)=1$ for any $x_0>0$. In addition, one
imposes the constraint $\lim\limits_{x_0\to\infty}S(t|x_0)=1$, since a
particle started from infinity cannot reach the origin in a finite
time. As a particle started from $x_0=0$ with a negative velocity
immediately hits the origin, one has $S(t|0)=0$. Finally, as we aim at
dealing with the second-order (in time) equation (\ref{eq:St_eq}), we
need to specify the initial value of the first time derivative, for
which we take $\partial_tS(0|x_0)=0$. This is a consequence of the
initial condition $S_\pm(0|x_0)=1$ and of the above backward
Fokker-Planck equation for $S_-(t|x_0)$.

The Laplace transform of Eq.~(\ref{eq:St_eq}) yields
\begin{eqnarray}
\nonumber
&&(p^2+2\lambda p-v^2\partial_{x_0}^2)\tilde{S}(p|x_0)\\
\nonumber
&&\quad=(2\lambda+p)S(0|x_0)+S'(0|x_0)=2\lambda+p\qquad(x_0>0),\\
\end{eqnarray}
where
\begin{equation}
\tilde{S}(p|x_0) = \int\limits_0^\infty dt \, e^{-pt} \, S(t|x_0)   \qquad (p > 0).
\end{equation}
Its solution on the positive semi-axis reads
\begin{equation}
\tilde{S}(p|x_0)=\frac{1}{p}-\frac{1}{p}e^{-\sqrt{p^2+2\lambda p}\, x_0/v},
\end{equation}
whereas its inverse Laplace transform is given by
\begin{eqnarray}
\label{eq:Stx}
S(t|x)&=&1-\Theta(t-x_0/v)\biggl[e^{-\lambda x_0/v}\\
\nonumber
&&+\lambda x_0/v\int\limits_{\lambda x_0/v}^{\lambda t}dz\frac{e^{-z}I_1(\sqrt{
z^2-(\lambda x_0/v)^2})}{\sqrt{z^2-(\lambda x_0/v)^2}}\biggr],
\end{eqnarray}
where $I_\nu(z)$ is the modified Bessel function of the first kind and $\Theta(
z)$ is the Heaviside step function: $\Theta(z)=1$ for $z>0$ and $0$ otherwise.
One sees that the survival probability is strictly unity for $t<x_0/v$, and
then it exhibits a jump $e^{-\lambda x_0/v}$, which corresponds to the
possibility to reach the target by a single displacement. The negative time
derivative of $S(t|x_0)$ gives the FPT-PDF
\begin{eqnarray}
\nonumber
H(t|x_0)&=&\delta(t-x_0/v)e^{-\lambda x_0/v}+\Theta(t-x_0/v)(x_0/v)\lambda\\
&&\times\frac{e^{-\lambda t}I_1(\lambda\sqrt{t^2-(x_0/v)^2})}{\sqrt{t^2-(x_0
/v)^2}}.
\label{eq:Htx}
\end{eqnarray}
The jump in the survival probability leads to a Dirac-$\delta$ distribution
in the FPT-PDF at $t=x_0/v$. At long times, we use $I_\nu(z)\simeq e^z/\sqrt{
2\pi z}$ to show that this PDF approaches the L{\'e}vy-Smirnov density with
the standard heavy tail,
\begin{equation}
\label{eq:Ht_inf}
H(t|x_0)\simeq\frac{x_0e^{-x_0^2/(4Dt)}}{\sqrt{4\pi Dt^3}}\propto t^{-3/2}
\qquad (t\to\infty),
\end{equation}
where $D=v^2/(2\lambda)$ is the effective diffusion coefficient. In particular,
the mean FPT is infinite, as for ordinary diffusion on the halfline.

It is convenient to express the above results in terms of the
parameters $\gamma= \lambda x_0/v$ and $\tmin=x_0/v$ by setting $\lambda=
v^2/(2D)=\gamma/\tmin$.  The survival probability from
Eq.~(\ref{eq:Stx}) then becomes
\begin{equation}
\label{eq:Stx2}
S(t|x_0)=1-\Theta(t-\tmin)\biggl[e^{-\gamma}+\gamma\hspace*{-2mm}\int
\limits_1^{t/\tmin}\hspace*{-2mm}dz\frac{e^{-\gamma z}I_1(\gamma
\sqrt{z^2-1})}{\sqrt{z^2-1}}\biggr].
\end{equation}
Since the survival probability vanishes at infinity, $S(\infty|x_0)=0$, we get
the identity 
\begin{equation}
\label{eq:identity}
\int\limits_1^{\infty}dz\frac{e^{-\gamma z}I_1(\gamma\sqrt{z^2-1})}{\sqrt{z^2-1}}
=\frac{1-e^{-\gamma}}{\gamma}\qquad(\gamma>0).
\end{equation}

\begin{figure}
\includegraphics[width=0.99\columnwidth]{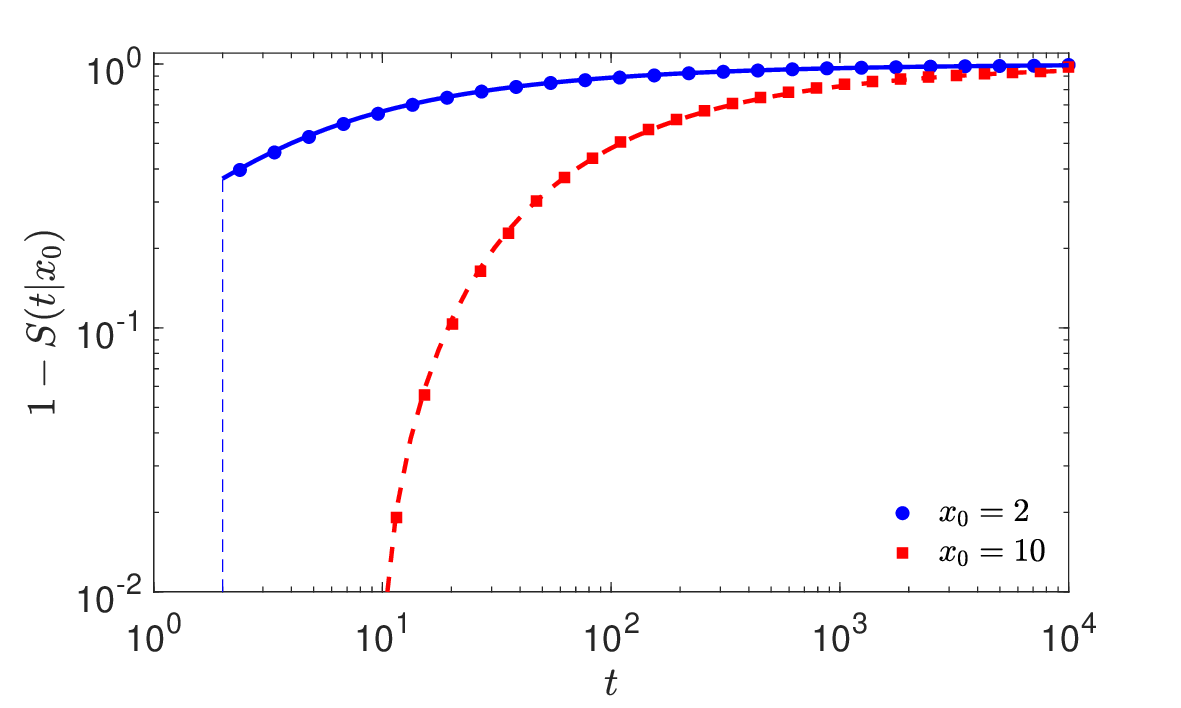} 
\caption{CDF of the FPT to the absorbing origin on the halfline, with $v=1$
and $\lambda=0.5$ such that $D=1$, and two values of $x_0$ as shown in the
caption. The solid line represents Eq.~(\ref{eq:Stx}), filled symbols show
the empirical CDF from Monte Carlo simulations with $M=10^4$ particles.}
\label{fig:St_halfline}
\end{figure}

Figure \ref{fig:St_halfline} illustrates the cumulative distribution function
(CDF) $1-S(t|x_0)$, either given by Eq.~(\ref{eq:Stx}), or estimated directly
from Monte Carlo simulations (see Sec.~\ref{sec:simu}). We observe excellent
agreement for both $x_0=2$ and $x_0=10$. Remarkably, the macroscopic
description provided by the telegrapher's equation, which a priori should
require a sufficiently large number of jumps (that corresponds to large $x_0$
and $t$), is so accurate even for moderate $x_0$.

\section{Fastest first-passage time}

This Section presents the mathematical details for our main results on the
fFPT $\T_N=\min\{\tau_1,\ldots,\tau_N\}$ among $N$ independent particles
started from the same point $x_0$. The probability law of this random variable
is
\begin{equation}
\label{eq:SNt}
\mathbb{P}_{x_0}\{\T_N>t \}=S_N(t|x_0)=[S(t|x_0)]^N,
\end{equation}
whereas its PDF is given by
\begin{equation}
\label{eq:HNt}
H_N(t|x_0)=NH(t|x_0)[S(t|x_0)]^{N-1},
\end{equation}
where $S(t|x_0)$ and $H(t|x_0)$ are from Eqs.~(\ref{eq:Stx}) and (\ref{eq:Htx}),
respectively.

\begin{figure}
\includegraphics[width=0.99\columnwidth]{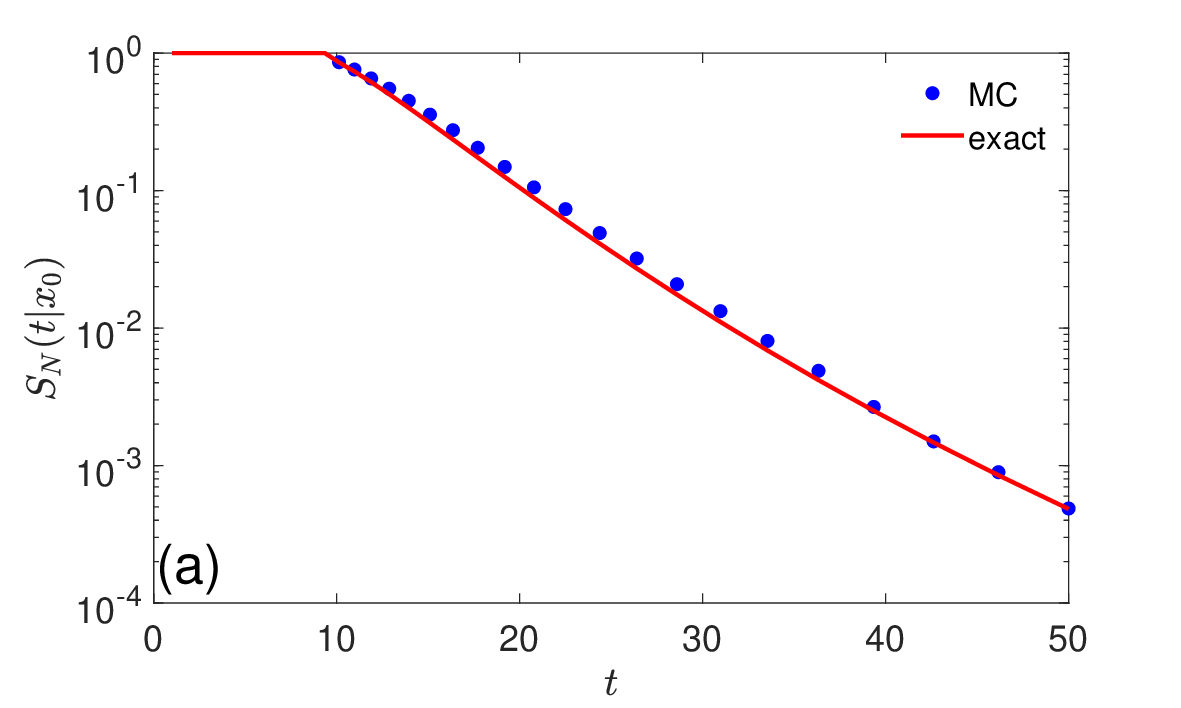} 
\includegraphics[width=0.99\columnwidth]{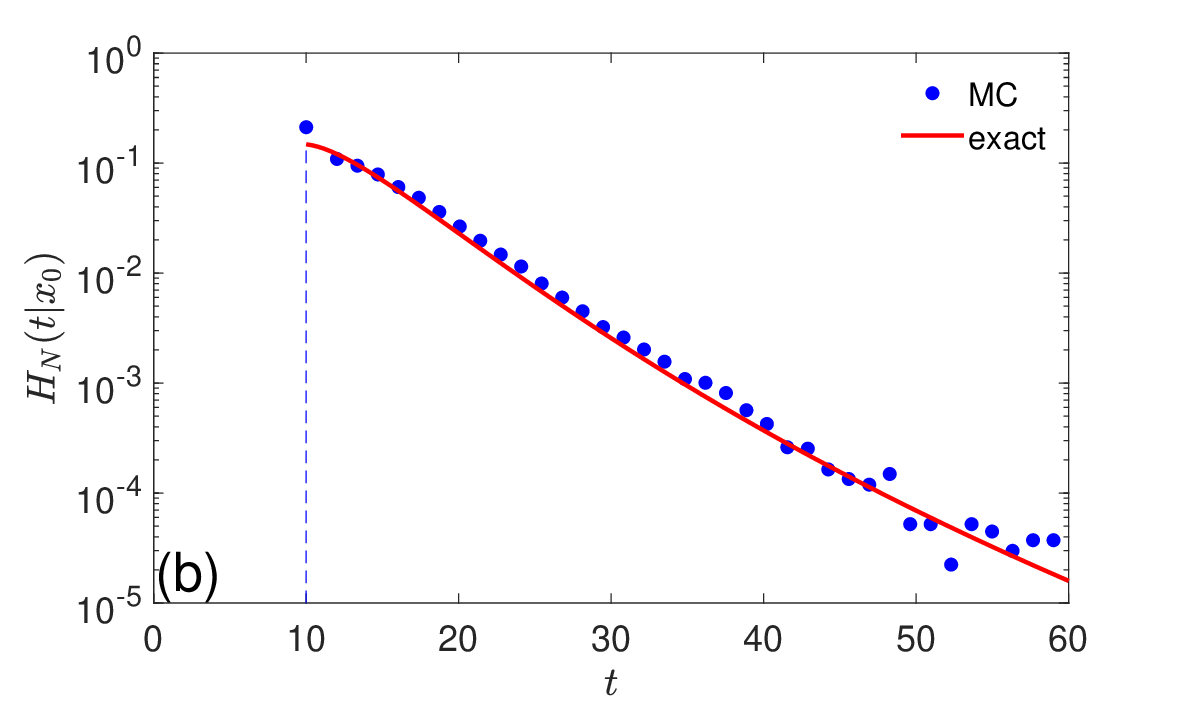} 
\caption{Survival probability {\bf (a)} and PDF {\bf (b)} of the fastest FPT
among $N=20$ particles on the halfline with absorbing origin, $v=1$, and
$\lambda=0.5$ such that $D=1$, and $x_0=10$. Solid lines show the exact
relations (\ref{eq:SNt}) and (\ref{eq:HNt}), with $S(t|x_0)$ and $H(t|x_0)$
given by Eqs.~(\ref{eq:Stx}) and (\ref{eq:Htx}), whereas filled circles
represent the empirical results from Monte Carlo simulations with $M=10^5$
particles.}
\label{fig:SNt_halfline}
\end{figure}

\begin{figure}
\includegraphics[width=0.99\columnwidth]{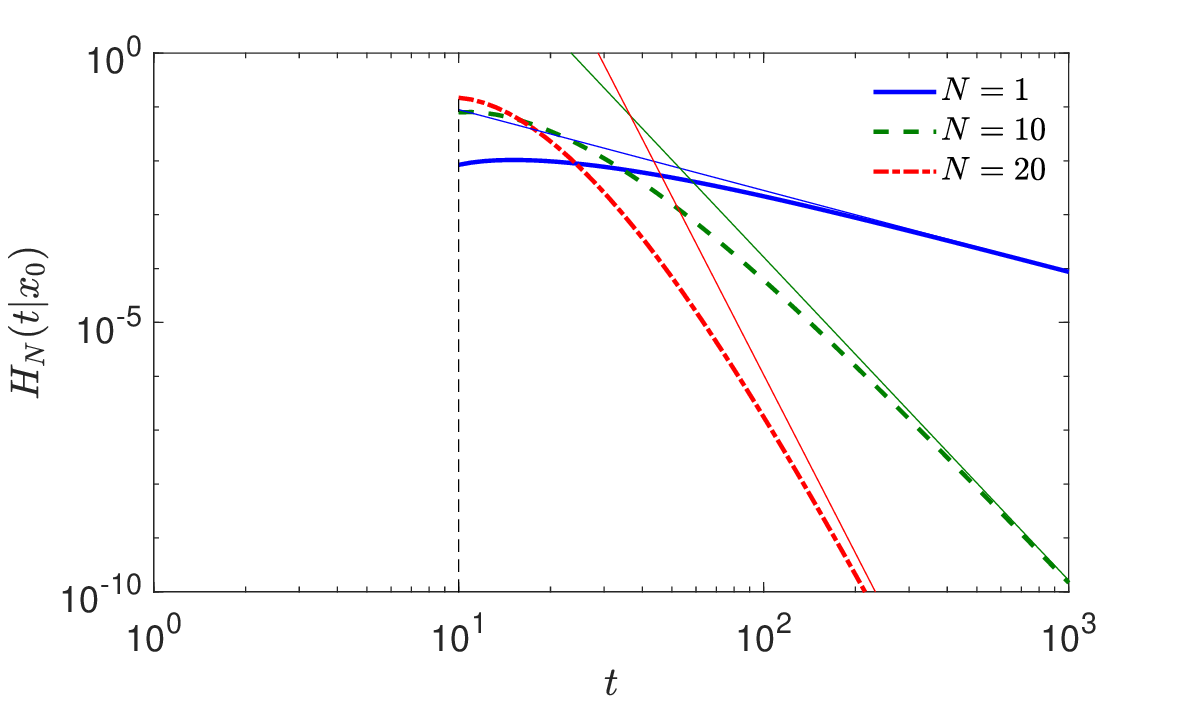} 
\caption{PDF $H_N(t|x_0)$ of the fFPT among $N$ particles on the halfline with
absorbing origin, $v=1$, and $\lambda=0.5$, such that $D=1$, and $x_0=10$.
Thick lines show the exact form (\ref{eq:HNt}), whereas thin lines represent
the long-time asymptotic behavior (\ref{eq:HNt_long}). The minimal value of the
fFPT is $\tmin=x_0/v=10$ (shown by the vertical dashed line), i.e., the PDF
is strictly $0$ for any $t<\tmin$.}
\label{fig:XNt_N}
\end{figure}

According to Eq.~(\ref{eq:Ht_inf}), one has $S(t|x)\simeq\sqrt{2T/(\pi t)}$ at
long times $t\gg T=x_0^2/(2D)$ such that
\begin{equation}
\label{eq:HNt_long}
H_N(t|x_0)\simeq N\frac{2^{3N/2-1}T^{N/2}}{\pi^{N/2}t^{1+N/2}}\qquad(t\to\infty).
\end{equation}
Panel (a) of Fig.~\ref{fig:SNt_halfline} compares the survival probability
$S_N(t|x_0)$ for $N=20$ independent particles with its empirical estimate,
whereas panel (b) shows the PDF. Figure \ref{fig:XNt_N} compares the PDFs
for the three values $N=1$, $10$, and $20$. In all cases, we observe the
long-time asymptotic behavior (\ref{eq:HNt_long}).

In the following, we focus on the mean fFPT and its variance and their
asymptotic behavior for large $N$.

\subsection{Mean fFPT}
\label{sec:Amean_fFPT}

Let us now consider the mean fFPT
\begin{equation}
\overline{\T_N}=\int\limits_0^\infty dt\,[S(t|x_0)]^N.
\end{equation}
According to the asymptotic behavior (\ref{eq:Ht_inf}), this integral diverges
for $N=1$ and $N=2$, as in the case of ordinary diffusion. The following
analysis is therefore restricted to $N\geq 3$.

The structure of $S(t|x)$ in Eq. (\ref{eq:Stx}) implies that
\begin{equation}
\label{eq:TN}
\overline{\T_N}=\frac{x_0}{v}\bigl(1+B_{N,\gamma}\bigr),  
\end{equation}
where
\begin{equation}
\label{eq:AN_def}
B_{N,\gamma}=\int\limits_1^\infty dy\,[f_\gamma(y)]^N,
\end{equation}
with
\begin{eqnarray}
\nonumber
f_\gamma(y)&=&1-e^{-\gamma}-\gamma\int\limits_1^y dz\frac{e^{-\gamma z}I_1(
\gamma\sqrt{z^2-1})}{\sqrt{z^2-1}}\\
&=&\gamma\int\limits_y^{\infty}dz\frac{e^{-\gamma z}I_1(\gamma\sqrt{z^2-1})}{
\sqrt{z^2-1}}\quad(y\geq1),
\label{eq:fgamma}
\end{eqnarray}
where we used Eq.~(\ref{eq:identity}) for the second equality. Changing the
integration variable $z=\cosh z'$, one gets Eq.~(\ref{eq:f_def}) in the main
text.

\subsection{Asymptotic behavior at large $N$}
\label{sec:Aasympt1}

In this Section, we inspect the asymptotic behavior of $B_{N,\gamma}$ in
the limit $N\to\infty$ for fixed $\gamma$. For this purpose, we rewrite
Eq.~(\ref{eq:AN_def}) in the form
\begin{equation}
B_{N,\gamma}=\int\limits_1^{\infty} dy \,e^{N\ln(f_\gamma(y))}.
\end{equation}
Since $\ln(f_\gamma(y))$ with $f_\gamma(y)$ defined in
Eq.~\eqref{eq:fgamma} is a monotonically decreasing function of $y$,
the above integral is supported by the behavior of $\ln(f_\gamma(y))$
in the close proximity of $y=1$. Using its Taylor expansion near
$y=1$,
\begin{equation}
\label{eq:lnfy}
\ln(f_\gamma(y))=\ln\left(1-e^{-\gamma}\right)-\frac{\gamma^2(y-1)}{2\left(
e^{\gamma}-1\right)}+\O\left((y-1)^2\right),
\end{equation}
we find to leading order in the limit $N\to\infty$ that
\begin{equation}
B_{N,\gamma}\approx\frac{2\left(e^{\gamma}-1\right)}{\gamma^2N}\left(1-e^{-
\gamma}\right)^N.
\end{equation}
Rewriting this expression formally we get our asymptotic result in
Eq.~\eqref{next}.  

One sees that the mean fFPT approaches a constant limit $x_0/v$ (the minimal
value of the fFPT), and this approach is {\it exponentially fast} with $N$:
$T_N-x_0/v\propto e^{-N/N_\gamma}$, where $N_\gamma$ is given by
Eq.~(\ref{eq:Ngamma}). In particular, in the limit of small $\gamma$, one
has $N_\gamma\approx1/\ln(1/\gamma)\to0$, i.e., the decay is very fast. In
the opposite limit of large $\gamma$, one gets $N_\gamma\approx e^{\gamma}
\to+\infty$, i.e., the decay becomes much slower, still being exponential
for any fixed $\gamma$. Since $N_\gamma$ grows exponentially fast with
$\gamma$, one may need to consider very large $N$ to observe the
exponential decay for large $\gamma$.

\subsection{Intermediate regime}
\label{sec:Aasympt2}

We seek an accurate uniform approximation of $f_\gamma(y)$ in
Eq.~\eqref{eq:f_def} for all $y$ away from the point $y=1$, which will permit
us to describe the behavior for $N$ in the interval $3 \leq N\ll N_{\gamma}$. To
this end, it is convenient to change the integration variable $z$ as $u=
\exp(-z)$, which leads to
\begin{equation} 
f_\gamma(y)=\gamma\int\limits_0^b\frac{du}{u}\exp\left(-\frac{\gamma}{2}\left(
\frac{1}{u}+u\right)\right)I_1\left(\frac{\gamma}{2}\left(\frac{1}{u}-u\right)
\right),
\end{equation} 
where $b=e^{-\arccosh(y)}=1/(y+\sqrt{y^2-1})$. The upper limit of integration
is less than unity, and the dominant contribution to the integral comes from
the region in the vicinity of $u=0$. Supposing that $\gamma$ is bounded away
from zero, we have
\begin{equation}
\frac{\gamma}{2}\left(\frac{1}{u}-u\right)\gg1, 
\end{equation}
which allows us to use the large-argument expansion for the modified Bessel
function of the first kind,
\begin{equation}
\label{approx}
I_1\left(\frac{\gamma}{2}\left(\frac{1}{u}-u\right)\right)\approx\sqrt{\frac{u}{
\pi\gamma}}\exp\left(\frac{\gamma}{2}\left(\frac{1}{u}-u\right)\right).
\end{equation}
This yields
\begin{equation}
\label{erf}
f_\gamma(y)\approx\sqrt{\gamma}\int_0^b\frac{du}{\sqrt{u}}\exp\left(-\frac{
\gamma}{2u}\right)=\erf\left(\sqrt{\gamma b}\right),
\end{equation}
where $\erf(x)$ is the error function. The approximation in
Eq.~\eqref{erf} describes $f_\gamma(y)$ very well for all $y$
sufficiently away from $y=1$, and the agreement improves as $\gamma$
increases. Indeed, at the \enquote{worst point} $y=1$, the exact value
is $f_\gamma(1)=1-e^{-\gamma}$, whereas Eq.~\eqref{erf} yields
$f_\gamma(1)=\erf(\sqrt{\gamma})$, which matches the exact result only
in the formal limit $\gamma=\infty$. In this regime, setting
$x=\sqrt{b}=(y+
\sqrt{y^2-1})^{-1/2}$ (and thus $y=(x^2+x^{-2})/2$), the relaxation function
$B_{N,\gamma}$ in Eq.~\eqref{Bf} can be written as
\begin{equation}
\label{tt}
B_{N,\gamma}\approx\int\limits^1_0dx\frac{1-x^4}{x^3}\bigl[\erf\left(\sqrt{
\gamma}x\right)\bigr]^N\qquad(N\geq 3),
\end{equation}
representing a very accurate approximation to the actual value of $B_{N,\gamma}$
for moderate $N$, such that $Ne^{-\gamma}<1$, for which the integral in
Eq.~\eqref{Bf} is supported on values of $y$ that remain away from $y=1$.

The integral on the right-hand-side of Eq.~\eqref{tt} can be estimated as
follows: a very tight upper bound on $\erf(x)$ for any value of $x\geq0$
(the error not exceeding $1$ percent) is given by \cite{will}
\begin{equation}
\erf(x)\leq\sqrt{1-\exp\left(-\frac{4x^2}{\pi}\right)}.
\end{equation}
Assuming for simplicity that $N$ is even, we bound the integral on the
right-hand-side of Eq.~\eqref{tt} from above as
\begin{eqnarray}
\nonumber
&&\hspace*{-0.6cm}\int\limits^1_0dx\frac{1-x^4}{x^3}\bigl[\erf\left(\sqrt{
\gamma}x\right)\bigr]^N\leq\lim\limits_{\epsilon\to0}\Biggl\{\sum_{j=0}^{N
/2}(-1)^j\binom{N/2}{j}\\
\nonumber
&&\qquad\qquad\times\int\limits^1_{\epsilon}dx\frac{1-x^4}{x^3}\exp\left(-\frac{
4\gamma j}{\pi}x^2\right)\Biggr\}\\
&&\hspace*{-0.6cm}=\lim\limits_{\epsilon\to0}\left[\frac{1}{2\epsilon^2}-1+
\frac{\epsilon^2}{2}+\sum_{j=1}^{N/2}(-1)^j\binom{N/2}{j}\Phi_{j}(\epsilon)
\right],
\end{eqnarray}
where
\begin{eqnarray}
\nonumber
\Phi_{j}(\epsilon)&=&\frac{\pi}{8 \gamma j}\left(\exp\left(-\frac{4\gamma j}{
\pi}\right)-\exp\left(-\frac{4\gamma j\epsilon^2}{\pi}\right)\right)\\
\nonumber
&&\hspace*{-0.8cm}-\frac{1}{2}\exp\left(-\frac{4\gamma j}{\pi}\right)+\frac{1}{
2\epsilon^2}\exp\left(-\frac{4\gamma j\epsilon^2}{\pi}\right)\\
&&\hspace*{-0.8cm}-\frac{2\gamma j}{\pi}{\rm Ei}\left(-\frac{4\gamma j}{\pi}
\right)+\frac{2\gamma j}{\pi}{\rm Ei}\left(-\frac{4\gamma j\epsilon^2}{\pi}
\right),
\end{eqnarray}
with ${\rm Ei}(-z)$ denoting the exponential integral
\begin{equation}
{\rm Ei}(-z)=-\int\limits^{\infty}_zdx\frac{\exp(-x)}{x}.
\end{equation}
Further on, we single out the part of $\Phi_j(\epsilon)$ dependent on $\epsilon$
and expand it into a Taylor series in powers of $\epsilon$. Summing it up over
$j$ we observe that the singular terms cancel each other when $N\geq 3$, as they
should---indeed, the integral in Eq.~\eqref{tt} exists for $\epsilon=0$ once
$N>2$. Assuming next that $\gamma$ is sufficiently large such that all terms
containing $\exp(-\gamma)$ can be safely neglected, as compared to the terms
containing only powers of $\gamma$, we get our estimate in Eq.~\eqref{inter},
where we used the following representation of the harmonic number
\begin{equation}
H_n=\sum_{j=1}^{n}\frac{(-1)^{j+1}}{j}\binom{n}{j}=\sum\limits_{j=1}^{n}
\frac{1}{j} ,
\end{equation} 
as well as the binomial identity
\begin{equation}
\sum\limits_{j=1}^{n}(-1)^jj\binom{n}{j}=0.
\end{equation}
Note that the harmonic number admits the large-$n$ expansion
\begin{equation}
H_n=\ln n+C+\frac{1}{2n}-\frac{1}{12n^2}+\O(n^{-4}),
\end{equation}
where $C\approx0.5772$ is the Euler constant.

To get the asymptotic behavior of the coefficients $S_n$ from Eq.~(\ref{eq:Sn}),
we first use the identity
\begin{equation}
\ln(j)=\int\limits_0^\infty du\frac{e^{-u}-e^{-ju}}{u}\quad(j\geq1)
\end{equation}
to deduce two equivalent integral representations of $S_n$, which are
suitable for its computation for large $n$,
\begin{equation} 
S_{n}=n \int\limits_0^\infty du\frac{e^{-u}(1-e^{-u})^{n-1}}{u}=n\int\limits_0
^1dz\frac{(1-z)^{n-1}}{-\ln z}.
\end{equation}
Setting $v=nz$ and expanding $1/(-\ln z)=1/(\ln n-\ln v)$ into a geometric
series, we get the asymptotic behavior of $S_n$ at large $n$,
\begin{equation}
S_n=\sum\limits_{k\geq0}\frac{m_k}{(\ln n)^k},\qquad m_k=\int\limits_0^\infty dv
e^{-v}(\ln v)^k.
\end{equation}
In particular, one has $m_0=1$, $m_1=-C$, $m_2=C^2+\pi^2/6$, etc.

\subsection{Diffusion limit}
\label{Sec:diffusion}

The diffusion limit corresponds to $v\to\infty$ and $\lambda\to\infty$ with $D
=v^2/(2\lambda)$ being fixed. In this case, one has $\gamma=x_0v/(2D)\to\infty$.

First, we check that the survival probability $S(t|x_0)$ approaches the expected
limit of ordinary diffusion as $\gamma\to\infty$. In fact, the Heaviside function
in Eq. (\ref{eq:Stx2}) can be replaced by unity, whereas $e^{-\gamma}$ vanishes.
Using the asymptotic behavior of the modified Bessel function, $I_1(z)\simeq e^z/
\sqrt{2\pi z}$ as $z\to\infty$, we can approximate the integral in
Eq.~(\ref{eq:Stx2}) to get
\begin{widetext}
\begin{eqnarray}
\nonumber
S(t|x_0)&\approx&1-\gamma\int\limits_{1}^{\gamma t/T}dz\frac{e^{-\gamma z+
\gamma\sqrt{z^2-1}}}{(z^2-1)^{3/4}\sqrt{2\pi\gamma}}\approx1-\frac{\sqrt{
\gamma}}{\sqrt{2\pi}}\int\limits_1^{\gamma t/T}dz\frac{e^{-\gamma/(2z)}}{(
z^2-1)^{3/4}}\\
\nonumber
&=&1-\frac{\sqrt{\gamma/2}}{\sqrt{\pi}}\int\limits_{2/\gamma}^{2t/T}dz\frac{
e^{-1/z}(\gamma/2)}{(\gamma^2z^2/4-1)^{3/4}}\approx1-\frac{\sqrt{\gamma/2}}{
\sqrt{\pi}}\int\limits_0^{2t/T}dz\frac{e^{-1/z} (\gamma/2)}{(\gamma/2)^{3/2}
z^{3/2}}\\
&=&1-\frac{1}{\sqrt{\pi}}\int\limits_0^{2t/T}dz\frac{e^{-1/z}}{z^{3/2}}=1-
\frac{2}{\sqrt{\pi}}\int\limits_{\sqrt{T/(2t)}}^\infty dz\, e^{-z^2}
=\erf(\sqrt{T/(2t)}),
\end{eqnarray}
\end{widetext}
where $T=x_0^2/(2D)$. In other words, we recover the expected diffusion limit,
\begin{equation}
S(t|x_0)\to\erf\bigl(x_0/\sqrt{4Dt}\bigr)\qquad(\gamma\to\infty).
\end{equation}

Next, we turn to the analysis of the mean fFPT. Substituting $v=2D\gamma/x_0$
into Eq.~(\ref{eq:TN}), we first rewrite it as
\begin{equation}
\label{eq:TN2}
\overline{\T_N}=T\frac{1+B_{N,\gamma}}{\gamma}.
\end{equation}
In order to evaluate the limit $\gamma\to\infty$ (for fixed $N$), we use the
approximate representation (\ref{tt}). Substituting $z=\sqrt{\gamma} x$, we
have
\begin{equation}
\frac{B_{N,\gamma}}{\gamma}\approx\int\limits_0^{\sqrt{\gamma}}\frac{dz}{z^3}
\bigl[\erf(z)\bigr]^N-\frac{1}{\gamma^2}\int\limits_0^{\sqrt{\gamma}}dzz\bigl[
\erf(z)\bigr]^N.
\end{equation}
The second integral vanishes as $\gamma\to\infty$, whereas the first integral
approaches the well-defined limit
\begin{equation}
\lim\limits_{\gamma\to\infty}\frac{B_{N,\gamma}}{\gamma}\approx\int\limits_0
^{\infty}\frac{dz}{z^3}\bigl[\erf(z)\bigr]^N.
\end{equation}
Its large-$N$ asymptotic behavior follows from the analysis of the mean fFPT
for ordinary diffusion,
\begin{equation}
\int\limits_0^\infty dt\bigl[\erf(x_0/\sqrt{4Dt})\bigr]^N=\frac{x_0^2}{2D}
\int\limits_0^\infty\frac{dz}{z^3}\bigl[\erf(z)\bigr]^N,
\end{equation}
so that
\begin{equation}
\int\limits_0^\infty\frac{dz}{z^3}\bigl[\erf(z)\bigr]^N\simeq\frac{1}{2\ln N}
+\O\bigl([\ln(N)]^{-2}\bigr).
\end{equation}
We therefore recover the leading-order term of the mean fFPT in the diffusion
limit.

\subsection{Variance and higher-order moments of the fFPT}
\label{sec:var}

According to Eq.~(\ref{eq:Stx2}), the $k$th-order moment of the fFPT is
\begin{eqnarray}
\nonumber
\overline{\T_N^k} &=& k\int\limits_0^\infty dt\,
t^{k-1}[S(t|x_0)]^N\\
&=&\tmin^k\biggl\{1+k\int\limits_1^\infty dy\, y^{k-1}[f_\gamma(y)]^N\biggr\},
\end{eqnarray}
where the function $f_\gamma(y)$ is defined in Eq.~(\ref{eq:fgamma}), and
$\tmin=x_0/v$. In particular, the variance reads
\begin{equation}
\label{eq:variance}
\frac{\mathrm{var}\{\T_N\}}{\tmin^2}=2\int\limits_1^\infty dy(y-1)
[f_\gamma(y)]^N-\biggl(\int\limits_1^\infty dy[f_\gamma(y)]^N\biggr)^2.
\end{equation}

In the limit $N\to\infty$ with fixed $\gamma$, one employs again the Taylor
expansion (\ref{eq:lnfy}) to get
\begin{equation}
\label{eq:variance_Ninf}
\frac{\mathrm{var}\{\T_N\}}{\tmin^2}\simeq(1-e^{-\gamma})^N
\frac{4(e^{\gamma}-1)^2}{\gamma^4 N^2}\bigl(2-(1-e^{-\gamma})^N\bigr),
\end{equation}
which is valid for $N\gg N_\gamma$. While the mean fFPT approaches its
minimal value $\tmin$, the variance vanishes exponentially fast, i.e., the
PDF of the fFPT becomes concentrated near $\tmin$.

In the intermediate regime $3 \leq N\ll N_\gamma$, one can adapt the asymptotic
technique from Sec.~\ref{sec:Aasympt2} to analyze the variance. However, we
skip this analysis and employ numerical computation of the integrals in
Eq.~(\ref{eq:variance}) to illustrate its behavior (see Fig.~\ref{fig:std_mean1}).

\section{Simulations for anomalous diffusion}
\label{sec:simu}

In this Section, we describe Monte Carlo simulations for anomalous diffusion
governed by dichotomous noise. As the particles move independently from each
other, we restrict the construction to a single particle trajectory, omitting
the index $k$ enumerating different particles.

The dichotomous noise process can be formally defined in terms of independent
identically distributed random variables $\delta_i$, obeying the exponential
law with the mean $1/\lambda$,
\begin{eqnarray}
\nonumber 
\eta(t)&=& v(-1)^{n(t)+n_0},\\
n(t)&=&\min\left\{n>0:\sum\nolimits_{i=1}^n\delta_i\geq t\right\},
\label{eq:eta_def}
\end{eqnarray}
where $n_0$ characterizes the choice of the sign in the first step (in our
simulations, we set $n_0=1$ to start with the negative sign). Substituting
Eq.~(\ref{eq:eta_def}) into Eq.~(\ref{eq:Langevin}), one gets
\begin{eqnarray}
\nonumber
x(t)&=&x_0+\frac{vT_0^{1-\beta}(-1)^{n_0}}{\Gamma(\beta+1)}\\
&&\hspace*{-0.6cm}\times\biggl\{t^\beta+2\sum\limits_{i=1}^{n(t)}(-1)^i\bigl(
t-(\delta_1+\ldots+\delta_i)\bigr)^\beta\biggr\},
\label{eq:xt}
\end{eqnarray}
where we introduced a shortcut notation $\beta=(\alpha+1)/2$.  For
practical purposes, we aim at evaluating $x(t)$ at times
$t_i=\delta_1+\ldots+\delta_i$. Note that $n(t_n)=n$ by construction
such that
\begin{eqnarray}
\nonumber
x(t_n)&=&x_0+\frac{vT_0^{1-\beta}(-1)^{n_0}}{\Gamma(\beta+1)}\biggl\{t_n^\beta
+2\sum\limits_{i=1}^{n-1}(-1)^i(t_n-t_i)^\beta\biggr\},\\
\label{eq:xtn}
x(t_1)&=&x_0+\frac{vT_0^{1-\beta}(-1)^{n_0}}{\Gamma(\beta+1)}t_1^\beta.
\end{eqnarray}
In the case of ordinary diffusion ($\alpha=1$), we have $\beta=1$ and obtain
\begin{equation}
x(t_n)=x(t_{n-1})+v\delta_n(-1)^{n-1+n_0}\quad(n=1,2,\ldots),
\end{equation}
with $x(0)=x_0$, as expected. This recurrence relation is used for
Monte Carlo simulations in the diffusive regime ($\alpha = 1$).  In
fact, for $N$ particles, we advance their positions in parallel by
generating waiting times. When one particle has crossed the target
(i.e., $x(t_n)<0$), the simulation is stopped, and the crossing time
is evaluated. We mention a potential drawback of this simulation: if
one particle is advanced by large $\delta_n$, another particle might
cross the origin earlier. However, as the probability of getting
abnormally large $\delta_n$ is exponentially small, the induced error
is small. We checked that it was actually negligible.

The situation is different for the anomalous case ($\alpha \ne 1$). In
fact, for each step $t_n$, one needs to re-evaluate the sum in
Eq.~(\ref{eq:xtn}) that would result in much slower simulations. Once
the simulation is stopped when $x(t_n)<0$, the crossing time $t_c$,
lying between $t_{n-1}$ and $t_n$, is evaluated by solving numerically
the equation $x(t_c)=0$, where $x(t)$ is given by Eq.~(\ref{eq:xt})
with $n(t)=n-1$.

It is worth noting that the minimal time to reach the target for a
single particle corresponds to a single move, so that
$x(t)=x_0-vT_0^{1-\beta} t^\beta/\Gamma(\beta+1)$, from which our
Eq. \eqref{eq:tmin_H} follows.  When $x_0/(vT_0)$ is large enough,
$\tmin$ is a monotonously decreasing function of $\beta$ (or
$\alpha$), as intuitively expected; in particular, superdiffusion
allows to reach the target {\it faster\/} than subdiffusion. However,
if $x_0/(vT_0)
\lesssim1$, the opposite trend occurs, i.e., $\tmin$ monotonously
increases with $\alpha$. Here we thus return to the counter-intuitive
situation when superdiffusion may look less efficient than
subdiffusion. We emphasize however, that this \enquote{paradox} is
rather artificial. In fact, when $\overline{x^2(t)}\sim t^{\alpha}$,
the MSD with a larger $\alpha$ grows faster only at long times; in
turn, it grows slower at short times.  As the first-passage time is
determined by short times, we are in the seemingly
\enquote{paradoxical} setting when superdiffusion is less efficient
than subdiffusion.


\begin{thebibliography}{40}

\bibitem{ptashne} M. Ptashne and A. Gann, {\it Genes and Signals} (Cold Spring Harbor
Laboratory Press, 2002).

\bibitem{alberts} B. Alberts, A. Johnson, J. Lewis, M. Raff, K. Roberts, and
P. Walter, {\it Molecular Biology of the Cell} (Garland Science, 2002).

\bibitem{olivier} O. B{\'e}nichou, C. Loverdo, M. Moreau, and R. Voituriez,
Intermittent search strategies, Rev. Mod. Phys. \textbf{83}, 81-130 (2011).

\bibitem{elf} P. Hammar, P. Leroy, A. Mahmutovic, E. G. Marklund, O. G. Berg,
and J. Elf, The lac repressor displays facilitated diffusion in living cells,
Science \textbf{336}, 1595-1598 (2012).

\bibitem{isaacsson} J. Ma, M. Do, M. A. Le Gros, C. S. Peskin, C. A. Larabell,
Y. Mori, and S. A. Isaacson, Strong intracellular signal inactivation produces
sharper and more robust signaling from cell membrane to nucleus, PLoS Comp.
Biol. \textbf{16}, e1008356 (2020).

\bibitem{maxplos} M. Bauer and R. Metzler, In vivo facilitated diffusion model,
PLoS ONE \textbf{8}, e53956 (2013).

\bibitem{snustad} D. P. Snustad and M. J. Simmons, {\it Principles of Genetics}
(Wiley, 2015).

\bibitem{target} D. S. Grebenkov, R. Metzler and G. Oshanin (Eds.) {\it Target Search
Problems} (Springer, Cham, CH, 2024).

\bibitem{Katja} G. H. Weiss, K. E. Shuler and K. Lindenberg, Order statistics
for first passage times in diffusion processes, J. Stat. Phys. {\bf 31},
255-278 (1983).

\bibitem{Meerson15} B. Meerson and S. Redner, Mortality, Redundancy, and
Diversity in Stochastic Search, Phys. Rev. Lett. {\bf 114}, 198101 (2015).

\bibitem{Reynaud15} K. Reynaud, Z. Schuss, N. Rouach, and D. Holcman,
Why so many sperm cells? Comm. Integr. Biol. {\bf 8}, e1017156 (2015).


\bibitem{Lawley20c} S. D. Lawley, Universal formula for extreme first passage
statistics of diffusion, Phys. Rev. E {\bf 101}, 012413 (2020).

\bibitem{Lawley20d} S. D. Lawley, Distribution of extreme first passage
times of diffusion, J. Math. Biol. {\bf 80}, 2301-2325 (2020).

\bibitem{Lawley20e} S. D. Lawley and J. B. Madrid, A Probabilistic Approach
to Extreme Statistics of Brownian Escape Times in Dimensions 1, 2, and 3,
J. Nonlinear. Sci. {\bf 30}, 1207-1227 (2020).

\bibitem{Madrid20} J. B. Madrid and S. D. Lawley, Competition between slow
and fast regimes for extreme first passage times of diffusion, J. Phys. A:
Math. Theor. {\bf 53}, 335002 (2020).

\bibitem{Schuss19} Z. Schuss, K. Basnayake, and D. Holcman, Redundancy
principle and the role of extreme statistics in molecular and cellular biology,
Phys. Life Rev. {\bf 28}, 52-79 (2019).


\bibitem{Grebenkov2020} D. S. Grebenkov, R. Metzler, and G. Oshanin, From
single-particle stochastic kinetics to macroscopic reaction rates: fastest
first-passage time of $N$ random walkers, New J. Phys. {\bf 22}, 103004 (2020).

\bibitem{adam} D. S. Grebenkov, R. Metzler, and G. Oshanin, Search
efficiency in the Adam-Delbr\"uck reduction-of-dimensionality scenario
versus direct diffusive search, New J. Phys. {\bf 24}, 083035 (2022).

\bibitem{aljazprx} A. Godec and R. Metzler, Universal proximity effect in
target search kinetics in the few encounter limit, Phys. Rev. X
\textbf{6}, 041037 (2016).

\bibitem{deniscomm} D. Grebenkov, R. Metzler, and G. Oshanin, Strong
defocusing of molecular reaction times: geometry and reaction control,
Comm. Chem. \textbf{1}, 96 (2018).

\bibitem{pccp} R. Metzler, J.-H. Jeon, A. G. Cherstvy, and E. Barkai,
Anomalous diffusion models and their properties: non-stationarity,
non-ergodicity, and ageing at the centenary of single particle tracking,
Phys. Chem. Chem. Phys. \textbf{16}, 24128 (2014).

\bibitem{Lawly2020} S. D. Lawley, Extreme statistics of anomalous subdiffusion
following a fractional Fokker-Planck equation: subdiffusion is faster than
normal diffusion, J. Phys. A: Math. Theor. {\bf 53}, 385005 (2020).

\bibitem{Hanggi95} P. H\"anggi and P. Jung, Colored noise in dynamical systems,
Adv. Chem. Phys. {\bf 89}, 239-326 (1995).

\bibitem{Bena06} I. Bena, Dichotomous Markov noise: exact results for
out-of-equilibrium systems, Int. J. Modern Phys. B {\bf 20}, 2825-2888 (2006).

\bibitem{Sandev22} T. Sandev, L. Kocarev, R. Metzler, and A. Chechkin,
Stochastic dynamics with multiplicative dichotomic noise: Heterogeneous
telegrapher's equation, anomalous crossovers and resetting, Chaos, Solitons
and Fractals {\bf 165}, 112878 (2022).

\bibitem{pastur} T. Herbeau, L. Pastur, P. Viot and G. Oshanin, Stochastic
gyration driven by dichotomous noises, J. Stat. Mech. \textbf{2026}, 013205
(2026).

\bibitem{cattaneo} C. R. Cattaneo, Sur une forme de l'{\'e}quation de la
chaleur {\'e}liminant le paradoxe d'une propagation instantan{\'e}e
(On a form of the heat equation eliminating the paradox of an instantaneous
propagation), C. R. Acad. Sci. Paris \textbf{247}, 431 (1958).

\bibitem{jou} D. Jou, J. Casas-V{\'a}zquez, and G. Lebon, {\it Extended
irreversible thermodynamics} (Springer, 2001).

\bibitem{we} O. B\'enichou, P. Illien, G. Oshanin, A. Sarracino, and
R. Voituriez, Tracer diffusion in crowded narrow channels, J. Phys.:
Cond. Matter {\bf 30}, 443001 (2018).

\bibitem{t2} J. Tailleur and M. E. Cates, Statistical mechanics of interacting
run-and-tumble bacteria, Phys. Rev. Lett. {\bf 100}, 218103 (2008).

\bibitem{t3} A. B. Slowman, M. R. Evans and R. A. Blythe, Jamming and
Attraction of Interacting Run-and-Tumble Random Walkers, Phys. Rev. Lett. {\bf
116}, 218101 (2016).

\bibitem{kurz} Y. Zhao, C. Kurzthaler, N. Zhou, J. Schwarz-Linek, C. Devailly,
J. Arlt, J.-D. Huang, W. C. K. Poon, T. Franosch, V. A. Martinez, and
J. Tailleur, Quantitative characterization of run-and-tumble statistics in
bulk bacterial suspensions, Phys. Rev. E {\bf 109}, 014612 (2024).


\bibitem{Dean21} D. S. Dean, S. M. Majumdar, and H. Schawe, Position
distribution in a generalized run-and-tumble process, Phys. Rev. E {\bf 103},
012130 (2021).

\bibitem{Malakar18} K. Malakar, V. Jemseena, A. Kundu, K. Kumar, S.
Sabhapandit, S. N. Majumdar, S. Redner, and A. Dhar, Steady state, relaxation
and first-passage properties of a  run-and-tumble particle in one-dimension,
J. Stat. Mech. \textbf{2018}, 043215 (2018).

\bibitem{Mori20} F. Mori, P. Le Doussal, S. N. Majumdar, and G. Schehr,
Universal survival probability for a d-dimensional run-and-tumble particle,
Phys. Rev. Lett. {\bf 124}, 090603 (2020).

\bibitem{urna} U. Basu, S. Sabhapandit, and I. Santra, Target Search by
Active Particles, in \cite{target}, pp. 463.

\bibitem{sm} Supplementary Information, \url{https://...}

\bibitem{protein_water} M. Yu, T. Castanheira Silva, A. van Opstal, S.
Romeijn, H. A. Every, W. Jiskoot, G.-J. Witkamp and M. Ottens, The
investigation of protein diffusion via H-cell microfluidics, Biophys. J.
{\bf 116}, 595-609 (2019).

\bibitem{ion_water} see, e.g., P. Swietach, M. Zaniboni, A. K. Stewart, A.
Rossini, K. W. Spitzer and R. D. Vaughan-Jones, 
Modelling intracellular H(+) ion diffusion, Prog. Biophys. Mol. Biol. {\bf 83}, 69-100 (2003).

\bibitem{timeref1} T. J. Lampo, S. Stylianidou, M. P. Backlund, P. A. Wiggins
and A. J. Spakowitz, Cytoplasmic RNA-Protein Particles Exhibit Non-Gaussian Subdiffusive Behavior,
Biophys. J. {\bf 112}, 532-542 (2017).

\bibitem{timeref2} C. Di Rienzo, V. Piazza, E. Gratton, F. Beltram and F.
Cardarelli, Probing short-range protein Brownian motion in the cytoplasm of
living cells, Nat. Comm. {\bf 5}, 5891 (2014). 

\bibitem{protein_cell} O. Seksek, J. Biwersi and A. S. Verkman,
Translational Diffusion of Macromolecule-sized Solutes in Cytoplasm
and Nucleus, J. Cell Biol.  {\bf 138}, 131-142 (1997).

\bibitem{will} J. D. Williams, An approximation to the probability integral,
Ann. Math. Statist. {\bf 17}, 363-365 (1946).

\bibitem{manzo} G. Mu{\~n}oz-Gil, G. Volpe, M. A. Garcia-March, E. Aghion, A.
Argun, C. B. Hong, et al.
Objective comparison of methods to decode anomalous diffusion, Nat. Comm.
\textbf{12}, 6253 (2021).

\bibitem{henrik} H. Seckler and R. Metzler, Bayesian deep learning for error
estimation in the analysis of anomalous diffusion, Nat. Comm. \textbf{13},
6717 (2022).

\bibitem{fbm} B. B. Mandelbrot and J. W. Van Ness, Fractional Brownian
motions, fractional noises and applications, SIAM Rev. \textbf{10}, 422-437 (1968).

\bibitem{mishura2008} Y. S. Mishura, \textit{Stochastic calculus for fractional
Brownian motion and related processes} (Springer-Verlag, Berlin, 2008).

\bibitem{vilk} O. Vilk, E. Aghion, T. Avgar, C. Beta, O. Nagel, A. Sabri, R.
Sarfati, D. K. Schwartz, M. Weiss, D. Krapf, R. Nathan, R. Metzler, and M.
Assaf, Unravelling the origins of anomalous diffusion: from molecules to
migrating storks, Phys. Rev. Res. \textbf{4}, 033055 (2022).

\bibitem{wei} W. Wang, Q. Wei, A. V. Chechkin, and R. Metzler, Different
behaviors of diffusing diffusivity dynamics based on three different
definitions of fractional Brownian motion, Phys. Rev. E \textbf{112}, 014108
(2025).

\bibitem{Lawley20b} S. D. Lawley, Extreme first-passage times for random walks
on networks, Phys. Rev. E {\bf 102}, 062118 (2020).



\end{thebibliography}
\end{document}